# RICH METHANE PREMIXED LAMINAR FLAMES

# DOPED BY LIGHT UNSATURATED HYDROCARBONS

# PART I: ALLENE AND PROPYNE

H.A.GUENICHE, P.A. GLAUDE, G. DAYMA, R. FOURNET, F.BATTIN-LECLERC*

Département de Chimie-Physique des Réactions,

Nancy University, CNRS, ENSIC,

1 rue Grandville, BP 20451, 54001 NANCY Cedex, France

Full-length article

SHORTENED RUNNING TITLE : **METHANE FLAMES DOPED BY ALLENE OR PROPYNE**

* E-mail : Frederique.battin-Leclerc@ensic.inpl-nancy.fr ; Tel.: 33 3 83 17 51 25 , Fax : 33 3 83 37 81 20

The structure of three laminar premixed rich flames have been investigated: a pure methane flame and two methane flames doped by allene and propyne, respectively. The gases of the three flames contain 20.9% (molar) of methane and 33.4 % of oxygen corresponding to an equivalent ratio of 1.25 for the pure methane flame. In both doped flames, 2.49 % of $C_3H_4$ was added, corresponding to a ratio $C_3H_4$ / $CH_4$ of 12 % and an equivalent ratio of 1.55. The three flames have been stabilized on a burner at a pressure of 6.7 kPa using argon as dilutant, with a gas velocity at the burner of 36 cm/s at 333 K. The concentration profiles of stable species were measured by gas chromatography after sampling with a quartz microprobe. Quantified species included carbon monoxide and dioxide, methane, oxygen, hydrogen, ethane, ethylene, acetylene, propyne, allene, propene, propane, 1,2-butadiene, 1,3-butadiene, 1-butene, iso-butene, 1-butyne, vinylacetylene and benzene. The temperature was measured using a thermocouple in PtRh (6%)-PtRh (30%) settled inside the enclosure and ranged from 700K close to the burner up to 1850K. In order to model these new results, some improvements have been made to a mechanism previously developed in our laboratory for the reactions of $C_3$-$C_4$ unsaturated hydrocarbons. The main reaction pathways of consumption of allene and propyne and of formation of $C_6$ aromatic species have been derived from flow rate analyses.

## INTRODUCTION

Soots and polyaromatic hydrocarbons (PAH), which are present in the exhaust gas of diesel engine, represent a high part of the urban pollution. Many efforts have then been focused on reducing the emissions of these compounds. The formation of soot precursors and PAH in combustion involves small unsaturated hydrocarbons, the chemistry of which is still very uncertain. Different reaction pathways have been proposed for the formation and the oxidation of the first aromatic compounds, involving the reactions of $C_2$ (acetylene), $C_3$ or $C_4$ unsaturated species [1-5].

As the determinant role of propargyl radicals in forming benzene, the first aromatic ring, is now well accepted, it is important to better understand their reactions. With that purpose, the oxidation of allene (propadiene) and propyne has been already studied in several oxidation conditions: shock tubes [6-8], flow reactor [9, 10], jet-stirred reactor [7, 11] and premixed flames [10, 11]. Previous works in flames include studies of the influence of the addition of allene on $H_2/O_2/Ar$ [12], $C_2H_2/O_2/Ar$ [13] and $C_2H_4/O_2/Ar$ [14] mixtures.

The purpose of the present paper is to experimentally investigate the structures of two premixed laminar methane flames containing propadiene and propyne, respectively, and to compare them with that of a pure methane flame containing the same mole fractions of methane and oxygen. The use of a methane flame will allow us to have a reactive mixture rich in methyl radicals and to be more representative of combustion mixtures containing larger hydrocarbons than hydrogen or $C_2$ flames. These results have been used to improve the mechanism previously developed in our laboratory for the reactions of $C_3$-$C_4$ unsaturated hydrocarbons (propyne, allene, 1-butyne, 2-butyne, 1,3-butadiene) [8, 15].

## EXPERIMENTAL PROCEDURE

The experiments were performed using an apparatus newly developed in our laboratory to study temperature and stable species profiles in a laminar premixed flat flame at low pressure. The body of the flat flame matrix burner, provided by McKenna Products, was made of stainless steel, with an outer diameter of 120 mm and a height of 60 mm (without gas/water connectors). This burner was built with a bronze disk (95% copper, 5% tin). The porous plate (60 mm diameter) for flame stabilization was water cooled (water temperature: 333 K) with a cooling coil sintered into the plate. The burner could be operated with an annular co-flow of argon to favor the stabilization of the flame.

This horizontal burner was housed in a water-cooled vacuum chamber evacuated by two primary pumps and maintained at 6.7 kPa by a regulation valve. This chamber was equipped of four quartz windows for an optical access, a microprobe for samples taking and a thermocouple for temperature measurements. The burner could be vertically translated, while the housing and its equipments were kept fixed. A sighting telescope measured the position of the burner relative to the probe or the thermocouple with an accuracy of 0.01 mm. The flame was lighted on using an electrical discharge. Gas flow rates were regulated by RDM 280 Alphagaz and Bronkhorst (El-Flow) mass flow regulators. The $C_3$ reactants (purity discussed in the text) and methane (99.95 % pure) were supplied by Alphagaz - L'Air Liquide. Oxygen (99.5% pure) and argon (99.995% pure) were supplied by Messer.

Temperature profiles were obtained using a PtRh (6%)-PtRh (30%) type B thermocouple (diameter 100 µm). The thermocouple wire was sustained by a fork and crosses the flame horizontally to avoid conduction heat losses. The junction was located at the centre of the burner. The thermocouple was coated with an inert layer of $BeO$-$Y_2O_3$ to prevent catalytic effects [16]. The ceramic layer was obtained by damping in a hot solution of $Y_2(CO_3)_3$ (93% mass.) and BeO (7% mass.) followed by a drying in a Mecker burner flame. This process was reiterated (about ten times) until the whole metal was covered. Radiative heat losses are corrected using the electric compensation method [17].

The sampling probe was in silica with a hole of about 50 μm diameter ($d_i$). The probe was finished by a small cone with an angle to the vertical of about 20°. For temperature measurements in the flames perturbed by the probe, the distance between the junction of the thermocouple and the end of the probe was taken equal to two times $d_i$, i.e. to about 100 μm. Gas samples were collected in a pyrex loop and directly obtained by connecting through a heated line the quartz probe to a volume, which was previously evacuated by a turbo molecular pump down to $10^{-7}$ kPa and which was then filled up to a pressure of 1.3 kPa (pressures were measured by a MKS 0-100 Torr pressure transducer). The pressure drop between the flame and the inlet of the probe ensured reactions to be frozen. Stable species profiles were determined by gas chromatography. Chromatographs with a Carbosphere packed column and helium or argon as carrier gas were used to analyse $O_2$, $H_2$, CO and $CO_2$ by thermal conductivity detection and $CH_4$, $C_2H_2$, $C_2H_4$, $C_2H_6$ by flame ionisation detection (FID). Heavier hydrocarbons (allene (a-$C_3H_4$), propyne (p-$C_3H_4$), cyclopropane (c-$C_3H_6$), propene ($C_3H_6$), propane ($C_3H_8$), vinylacetylene ($C_4H_4$), butadienes (1,2-$C_4H_6$, 1,3-$C_4H_6$), 1-butyne (1-$C_4H_6$), 1-butene (1-$C_4H_8$), iso-butene (i-$C_4H_8$) and benzene ($C_6H_6$)) were analysed on a Haysep packed column by FID and nitrogen as gas carrier gas. The identification of these compounds was performed using GC/MS and by comparison of retention times when injecting the product alone in gas phase. Figure 1 presents a typical chromatogram of $C_3$-$C_6$ compounds obtained for the flame doped with allene. Contrary to what is observed due to the scale of the figure, the separation between the peaks of iso-butene, 1-butene and 1,3-butadiene and those of vinylacetylene and 1-butyne is acceptable. The peaks of propane and cyclopropane cannot be distinguished. Water and small oxygenated compounds, such as acetone, acetaldehyde and formaldehyde, were detected by GC-MS but not quantitatively analysed. Toluene could be quantified for flames of richer mixtures, but was not detected in the flames studied here due to our limit of detection (around $1x10^{-6}$ in mole fraction).

FIGURE 1

Calibrations were performed by analysing a range of samples containing known pressures of each pure compound to quantify and mole fractions were derived from the known total pressure in the sampling line. Error on mole fractions was around ± 10% for major compounds, but can be more for minor ones.

## EXPERIMENTAL RESULTS

Three flames stabilized on the burner at 6.7 kPa with a gas flow rate of 3.29 l/min corresponding to a gas velocity at the burner of 36 cm/s at 333 K and mixtures containing 20.9% methane and 33.4% oxygen have been investigated:

- A pure methane flame supplied with a mixture containing also 45.6% argon and corresponding to an equivalent ratio of 1.25.
- Two doped flames supplied with mixtures containing also 43.2% argon and 2.49 % of allene or propyne and corresponding to an equivalent ratio of 1.55. $C_3$ compounds represented then 12% of the amount of methane. Chromatographic analysis showed that allene was 98.4% pure, containing propyne (1.2%), propene (0.25%) cyclopropane (0.08%), 1-butene (0.04%), 1,3-butadiene (0.04%), 1,2-butadiene (0.01%) and 1-butyne (0.04%) as major impurities. Propyne was 98.3% pure, but contained also allene (0.44%), propene (0.04%) cyclopropane (0.05%), 1-butene (0.6%), 1,3-butadiene (0.2%), 1,2-butadiene (0.3%) and 1-butyne (0.05%) as major impurities. The presence of cyclopropane as an impurity was mentioned by the provider.

Figure 2a presents the temperature profiles measured for the three flames without the probe. The results are similar for both doped flames and different from those in the unseeded flame due to the difference in equivalence ratios. The lowest temperatures measured the closest to the burner are around 700 K. In the pure methane flame, the highest temperatures are reached between 0.4 and 0.6 cm above the burner and are around 1850K. For the doped flames, temperature peaks between 0.75 and 0.95 cm above the burner at around 1880K and decreases thereafter because of heat losses.

Figure 2b displays the experimental profiles obtained with and without the probe in the case of the flame doped with allene and shows that the presence of the probe induces a thermal perturbation, as it reduces the temperature of more than 100K in some places.

FIGURE 2

Figures 3 and 4 present the profiles of the main species involved in the combustion of methane vs. the height above the burner for the three flames. Figure 3 displays the profiles of methane (fig. 3a) and oxygen (fig. 3b) and shows that the consumption of these species occurs further in the flames containing of allene and propyne. A displacement of the position of the maximum concentrations is also observed for carbon monoxide and $C_2$ compounds as shown in figures 3c and 4. The profile of carbon dioxide (fig. 3d) shows a marked inflexion point in the doped flames, which cannot be observed in the pure methane flame. Ethane (fig. 4d) is produced promptly in the three flames and reaches its maximum concentration close to the burner, around 0.2 cm in the pure methane flame and around 0.3 cm in the doped flames. The profile of ethylene (fig. 4c) peaks around 0.3 cm in the pure methane flame and around 0.4 cm in the doped flames, that of acetylene (fig. 4b) around 0.35 cm in the pure methane flame and around 0.5 cm in the doped flames and that of carbon monoxide (fig. 3c) around 0.45 cm in the pure methane flame and around 0.6 cm in the doped flames. While the maximum concentration of acetylene is strongly increased by the addition of allene and still more by that of propyne, the maximum concentrations reached for carbon oxides (fig. 3c and 3d) hydrogen (fig. 4a) and ethylene are changed by a more limited factor and that of ethane is almost unmodified.

FIGURES 3 AND 4

Figure 5 presents the profiles of the observed $C_3$ products and shows that very small amounts of these products are also observed in the pure methane flame. The consumptions of allene and propyne (fig. 5a), as reactant, are the same in both doped flames. The most important $C_3$ products are allene in the propyne flame and propyne in the allene flame (fig. 5b). An important formation of propene (fig. 5c) is also encountered. The maximum concentrations of these $C_3$ products are located at around 0.2 cm from the burner. The fact that larger amounts of propyne and propene are observed in the allene flame in comparison with allene and propene in the propyne flame, as displayed in figures 5b and 5c, is partially due to the presence of larger concentrations of propyne (1.2%) and propene (0.25%) as an impurity in allene than it is the case for allene (0.4%) and propene (0.04%) in propyne. As it is observed in similar amounts in the three flames, we have considered that the product shown on figure 5d, which has the same retention time as propane and cyclopropane, is mainly propane.

FIGURE 5

Figure 5 also displays the profiles of benzene (fig. 5e), which present a maximum at around 0.25 cm from the burner. It shows that a larger maximum concentration of benzene is obtained in the propyne flame than in the allene flame (around a factor 2.5); this difference between both $C_3$ hydrocarbons, as that concerning acetylene, is in agreement with previous experimental results obtained in a jet stirred reactor operating between 800 and 1200 K, at 10 atm with stoichiometric mixtures [11].

Figure 6 presents the profiles of $C_4$ products and shows that 1-butene (fig. 6c), iso-butene (fig. 6e) and 1-butyne (fig. 6d) are formed close to the burner with maximum concentrations at around 0.2 cm, while 1,3-butadiene (fig. 6a), 1,2-butadiene (fig. 6b) and vinylacetylene (fig. 6f) are formed further in the flame. While the formations of 1-butyne and vinylacetylene are similar in both doped flames and that of iso-butene is larger in the allene flame, it is difficult to conclude for 1-butene, 1,3-butadiene and 1,2-butadiene, which, as stated previously, are present as an impurity in more important amounts in propyne than in allene.

FIGURE 6

Benzene and C4 compounds were not detected in the pure methane flame.

## DESCRIPTION OF THE PROPOSED MECHANISM

This mechanism is an improvement of our previous mechanism that was built to model the oxidation of $C_3$-$C_4$ unsaturated hydrocarbons [8, 15] to better take into account the reactions of allene, propyne and propargyl radicals.

*Reaction base for the oxidation of $C_3$-$C_4$ unsaturated hydrocarbons [8]*

This $C_3$-$C_4$ reaction base was built from a review of the recent literature and is an extension of our previous $C_0$-$C_2$ reaction base [18]. This $C_0$-$C_2$ reaction base includes all the unimolecular or bimolecular reactions involving radicals or molecules including carbon, hydrogen and oxygen atoms and containing less than three carbon atoms. The kinetic data used in this base were taken from the literature and are mainly those proposed by Baulch *et al.* [19] and Tsang *et al.* [20]. The $C_0$-$C_2$ reaction base was first presented in the paper of Barbé *et al.* [18] and has been up-dated [8].

The $C_3$-$C_4$ reaction base includes reactions involving $C_3H_2$ ($CH{\equiv}CCH{\bullet\bullet}$), $C_3H_3$ ($CH{\equiv}CCH_2{\bullet} \leftrightarrow {\bullet}CH{=}C{=}CH_2$), $C_3H_4$ (allene and propyne), $C_3H_5$ (3 isomers ($aC_3H_5$: ${\bullet}CH_2CH{=}CH_2$, $sC_3H_5$: $CH_3CH{=}CH{\bullet}$, $tC_3H_5$: $CH_3C{\bullet}{=}CH_2$)), $C_3H_6$, $C_4H_2$, $C_4H_3$ (2 isomers ($nC_4H_3$: ${\bullet}CH{=}CHC{\equiv}CH$, $iC_4H_3$: $CH_2{=}C{\bullet}C{\equiv}CH \leftrightarrow CH_2{=}C{=}C{=}CH{\bullet}$)), $C_4H_4$, $C_4H_5$ (5 isomers ($nC_4H_5$: ${\bullet}CH{=}CHCH{=}CH_2$, $iC_4H_5$: $CH_2{=}CHC{\bullet}{=}CH_2 \leftrightarrow {\bullet}CH_2CH{=}C{=}CH_2$, $C_4H_5$-1s: $CH_3CH{\bullet}C{\equiv}CH$, $C_4H_5$-1p: $CH_2{\bullet}CH_2C{\equiv}CH$, $C_4H_5$-2: $CH_3{\bullet}C{\equiv}CCH_2$)), $C_4H_6$ (1,3-butadiene, 1,2-butadiene, methyl-cyclopropene, 1-butyne and 2-butyne), as well as the formation of benzene. Pressure-dependent rate constants follow the formalism proposed by Troe [21] and efficiency coefficients have been included, when the related data were available. This reaction base was built in order to model experimental results obtained in a jet-stirred reactor for methane and ethane [18], profiles in laminar flames of methane, acetylene and 1,3-butadiene [8] and shock tube auto-ignition delay times for acetylene, propyne, allene, 1,3-butadiene [8], 1-butyne and 2-butyne [15].

Thermochemical data are estimated by the software THERGAS developed in our laboratory [22], which is based on the additivity methods proposed by Benson [23].

*Reactions related to allene, propyne, propargyl radicals and related $C_4$ species*

The part of mechanism, described below and given in Table I, is included in a file which also contains the two reactions bases described above and which can be used to run simulations using CHEMKIN [24]. In order to correctly model the consumption of benzene, our recent primary and secondary mechanisms for the oxidation of this species [25] should also been added. A short mechanism of propane, 1-butene and iso-butene has also been considered based on our work on alkanes and alkenes [26], as well as four reactions of cyclopropane (isomerisation to propene, H-abstraction to give allyl radical). The whole mechanism, which is available on request, includes 812 reactions involving 115 species.

**TABLE I**

For both propyne and allene, the pressure dependence of the rate constants of the additions of H-atoms to give $C_3H_5$ radicals (reactions 6, 7, 27, 28 in Table 1) has been taken into account by considering high and low pressure limits. We have reevaluted the rate constants of the bimolecular initiations with oxygen molecule based on the work of Ingham *et al.* [39]. For allene (reaction 14), A was taken equal to 4 x $7.10^{12}$ $cm^3.mol^{-1}s^{-1}$, as there are four abstractable vinylic hydrogen atoms, while for propyne (reaction 33), A was taken equal to 3 x $7.10^{11}$ $cm^3.mol^{-1}s^{-1}$, as there are three abstractable allylic hydrogen atoms; for both compounds, the activation energy was set equal to the enthalpy of reaction.

The main addition to the reactions of allene was to consider the reactions of abstraction of a hydrogen atom by H atoms and OH, $CH_3$, $C_2H$, $C_2H_3$, $C_2H_5$ and a-$C_3H_5$ radicals (reactions 15-22). These reactions, involving the abstraction of vinylic H-atoms, had been neglected in our previous work in shock tube conditions [8], but were found important here to correctly reproduce the profile of allene in the flame doped with this compound. For the rate constants, which were not available in the literature, we have used the same value as for the abstraction of hydrogen atoms from 1,3-butadiene to give the resonance stabilized i$C_4H_5$ radicals [8].

Concerning allene, the isomerization to propyne (reaction 1) has been written reversible and the bimolecular initiation between two allene molecules to give allyl and propargyl radicals (reaction 3) has been taken into account as proposed by Dagaut *et al* [28]. We have also considered the additions of methyl radicals to the double bonds of allene to give 1-buten-2-yl (reaction 12) and iso-butenyl (reaction 13) radicals. According to Aleksandrov *et al* [30], the products of the addition of oxygen atoms to allene (reaction 9) are hydrogen atoms and $C_2H_3CO$ radicals which decomposed rapidly to give carbon monoxide and vinyl radicals, instead of ethylene and carbon monoxide, as previously written [8].

The unimolecular initiation of propyne to give $C_2H$ and $CH_3$ radicals (reaction 25) has been revisited using the software KINGAS [40], because the activation energy proposed by Wu and Kern (100 kcal/mol [32]) was too weak compared to the enthalpy of reaction (125 kcal/mol). To improve our modeling of the formation of acetylene, we have considered the addition of H-atoms to propyne to give acetylene and methyl radicals (reaction 26) in a single step with the rate constant proposed by Hidaka *et al.* [27]. A second channel giving acroleine and H-atoms has been considered for the addition of OH radicals to propyne (reaction 31). For the abstraction of hydrogen atoms from propyne by oxygen atoms (reaction 39), we have used the rate constant calculated by Adusei *et al.* [35].

Concerning propargyl radicals, we have added the decomposition to give H-atoms and $C_3H_2$ (reaction 42) and the combination with $HO_2$ radicals (reaction 50). In the case of the reactions leading to aromatic compounds, we have kept, the previous value of $1.10^{12}$ $cm^3.mol^{-1}s^{-1}$ for the recombination of two propargyl radicals to give phenyl radicals and H-atoms (reaction 51), as it is in good agreement with what has been recently proposed by Miller *et al.* [42] and Rasmussen *et al.* [43]. We have removed from our previous mechanism the addition of propargyl radicals to allene leading to benzene and H-atoms and the recombination between allene and propargyl radicals to produce phenyl radicals and two H-atoms, which were not considered by Rasmussen *et al.* [43] and which induce an overprediction of benzene in the case of the flame doped with allene.

## COMPARISON BETWEEN EXPERIMENTAL AND SIMULATED RESULTS

Simulations were performed using PREMIX from CHEMKIN II [24]. The presence of hydrocarbon impurities in the $C_3$ reactants was taken into account in simulations; i.e. the reactants in our simulations files have the $C_3$ and $C_4$ hydrocarbons composition mentioned in page 6. To compensate the thermal perturbations induced by the presence of the quartz probe and the thermocouple in the reactive mixture, the temperature profile used in calculations is an average between the experimental profiles measured with and without the quartz probe, shifted 0.1 cm away from the burner surface, as shown in figure 2b in the case of the flame doped with allene.

Figures 3 and 4 show that the model reproduces satisfactorily the consumption of reactants and the formation of the main products related to the consumption of methane in the case of the three flames. The slower consumption of methane and oxygen when added a $C_3$ compound is well captured, as well as the similar reactivity in the two doped flames. Calculations predict well the increase in the formation of acetylene when added allene and the still higher one when added propyne. To decouple the effect due to the increase of equivalence ratio (Φ) and that induced by the presence of a $C_3$ unsaturated compound, these figures display also the results of a simulation performed for a flame containing 20.9% methane and 27.6% oxygen (with no $C_3$ additive) for Φ= 1.55, i.e. equal to that of the doped flames. As the temperature rise is mainly influenced by Φ, we have used the same temperature profile as to model the flame containing allene. This simulation shows that the experimental differences encountered for the profiles of methane, oxygen, ethane and ethylene when added allene or propyne are only due to the increase of Φ (the profiles can almost not be distinguished from those obtained in the flame doped with allene). The profiles of carbon oxides and hydrogen are also modified by the change of Φ, but they are different in the doped and pure methane flames at Φ=1.55 due to a difference in the C/O and C/H ratios. For Φ= 1.55, the profile of carbon dioxide in a pure methane flame presents also a marked inflexion points, showing that this effect is mainly related to equivalence ratio, i.e. in a rich mixture, carbon dioxide exhibits a more marked secondary behavior due to the fact that the formation of carbon monoxide is favored. In the case of acetylene, the increase between pure methane and doped flames is a factor 4 for allene and 5 for propyne, while the difference of Φ would only explain a factor around 3. There is then a specific way of formation of this species involving our $C_3$ additives.

Figure 5 shows that the consumption of the $C_3$ reactants in the doped flames is well predicted and that the peak of the profile of the concentration of the isomer of the reactant, propene and propane in both doped flames are captured within a factor better than 2. Simulations show that the formation of cyclopropane is negligible.

The simulated profiles of $C_4$ compounds are displayed in Figure 6. The prediction is better than a factor 3 for butadienes, 1-butyne and 1-butene in both seeded flames, which is acceptable taking into account the errors in the analyses of these minor species and the uncertainties in the rate constants used in our model. Despite that butadienes are present as an impurity in the $C_3$ reactant, simulations clearly show a formation of these compounds in the seeded flames. Simulations overpredict the formation of vinylacetylene by a factor 5 in both doped flames, but reproduce well that the same amount is obtained in both doped flames. The production of iso-butene, which is a very minor product, is also badly reproduced in both doped flames.

The model reproduces well the formation of benzene in the two doped flames and the differences observed between both $C_3$ additives, especially.

**DISCUSSION**

Figure 7 displays the flows of consumption of the $C_3$ reactant in the two doped flames at a temperature about 720 K corresponding to a 35% conversion. The two main pathways of consumption for both compounds are additions and H-abstractions to give propargyl radicals, H-abstractions accounting for 20% of the consumption of allene and for 40% of that of propyne. A faster production of propargyl radicals in the case of propyne, which is due to the lower activation energies of H-abstractions, can be slightly seen on the simulated profiles of these radicals displayed in fig. 8a, even if this effect is compensated by a more rapid consumption of these radicals. At low temperature, in the zone close to the burner, resonance stabilized propargyl radicals are mainly consumed by combinations with H atoms to give back the initial reactant or its isomer, with $HO_2$ radicals to give formaldehyde and $C_2H$ and OH radicals, with methyl radicals to produce 1-butyne or 1,2-butadiene or with themselves to form H atoms and phenyl radicals. At higher temperature, further in the flame, the combinations with H-atoms and $HO_2$ radicals become less favoured and propargyl radicals are less consumed.

Figure 8b displays the profiles of hydrogen atoms in the two doped flame and indicates that the pool of small radicals is slightly lower in the allene flame than in the propyne flame. This effect is partially due to the fact that addition of small radicals are favored in the case of allene which includes two double bonds. That explains why the integrated amount of propargyl radicals is larger in the allene flame, in which they are less consumed by combinations, than in the propyne flame, as shown in fig. 8a. This difference in the consumption of propargyl radicals explains partly why larger amounts of benzene are obtained in the allene flame as it can be seen in fig. 5e. Another consequence of the prompter formation and consumption of propargyl radicals in the propyne flame is that the formation of phenyl radicals occurs in a mixture rich in oxygen and then more favorable to the oxidation of aromatic radicals (the only reactions under the conditions of fig. 7) than to their reaction with H-atoms to give benzene.

Under the conditions of fig. 7, the branching ratio of the combination of H atoms with propargyl radicals is more than 90 % towards the formation of propyne, that explains why the formation of propyne in the allene flame is larger than that of allene in the propyne flame as shown in fig. 5b. Other reactions of propargyl radicals include reaction with oxygen molecules to give ketene and HCO radicals, and H-abstraction by H atoms and OH radicals to give $C_3H_2$, which react also with oxygen molecules. 1,3-butadiene is obtained from 1,2-butadiene. Vinyl acetylene is a degradation product from 1-butyne and 1,2-butadiene and uncertainties in the kinetics of the reactions consuming these species could explain a deteriorated prediction.

FIGURES 7 AND 8

The main additions to allene or propyne are that of OH radicals and that of H atoms. The additions of OH radicals lead to oxygenated products and account for 26 % of the consumption of allene and to 13% of that of propyne. The additions of H atoms to allene lead to t-$C_3H_5$ radicals (40% of the consumption), which react rapidly with oxygen molecules to give carbon monoxide, formaldehyde and methyl radicals, and to allyl radicals (14% of the consumption). The resonance stabilized allyl radicals react mainly by combination with H atoms to produce propene, with $HO_2$ radicals to give acroleine, H atoms and OH radicals and with methyl radicals to produce 1-butene. Taking into account that the rate constants concerning allyl radicals are amongst the best established in our model [37], it is surprising that the formation of propene is not better estimated by our model in the allene flame. In the propyne flame, 30 % of propene is obtained by reaction of propyl radicals with oxygen molecules. Propyl radicals derive from propane, which is produced by recombination of methyl and ethyl radicals. The addition of H atoms to propyne leads to methyl radical and acetylene and explains why the maximum of the profile of this last species is higher in the propyne flame than in the allene flame. In the allene flame, acetylene is 50% obtained from the addition of H atoms to propyne, which is formed by combination of H atoms and propargyl radicals, and 50% obtained from reactions of $C_2H$ (deriving from propargyl radicals) and vinyl radicals (deriving from acroleine). Very minor channels involve the addition of methyl radicals to allene to form iso-butyl and 2-butyl radicals, which lead to iso-butene and 1-butene, respectively. Cyclopropane is consumed to give allyl radicals, but there is no production channel, even if the isomerisation from propene has been considered in our mechanism.

Figure 9 displays the reaction flows of phenyl radicals in the flame seeded with allene at the position of the maximum of the benzene profile, corresponding to a temperature about 1290 K and a 75% conversion of allene. Under these conditions, phenyl radicals are almost exclusively formed by the combination of propargyl radicals (the H abstractions from benzene are still slow) and only 4% of them are consumed by combination with H atoms to form benzene. The main consumption of phenyl radicals are due to reactions with oxygen molecules to give O atoms and phenoxy radicals or H atoms and benzoquinone, which decomposes to form cyclopentadienone and carbon monoxide. Resonance stabilized phenoxy radicals react by combination with H atoms to give phenol or by decomposition to give carbon monoxide and cyclopentadienyl radicals and then cyclopentadiene, which was certainly present in too low amount to be detected in our $C_3$-$C_6$ analysis.

FIGURE 9

## CONCLUSION

This paper presents new experimental results for rich premixed laminar flames of methane seeded or not with allene and propyne, as well as some improvements made to the mechanism previously developed in our laboratory for the reactions of $C_3$-$C_4$ unsaturated hydrocarbons. Profiles of temperature have been measured and profiles of stable species have been obtained for 17 products, including benzene and six $C_4$ unsaturated compounds that are soot precursors. The use of methane as the background and consequently of a flame rich in methyl radicals favors the formation of $C_4$ compounds from the $C_3$ compounds.

The presence of $C_3$ additives strongly promotes the formation of acetylene, which can be mainly attributed to the addition of hydrogen atoms to propyne. The presence of allene and propyne is also responsible for the formation of benzene, which cannot be detected in the pure methane flame. This formation is totally due to the recombination of propargyl radicals and it is worth noting that simulation using a value of the rate constant close to that now admitted in the literature leads to a good agreement.

Many similarities exist between allene and propyne flames in both reactivity and products formation. Nevertheless, the pool of small radicals is slightly larger in the propyne flame, which leads to a faster consumption of propargyl radicals and a lower formation of benzene than in the flame seeded by allene.

In line with this work, studies of flames doped with 1,3-butadiene and cyclopentene are in progress using the same methodology.

## ACKNOWLEDGEMENTS

The authors thank J.F. Pauwels and P. Desgroux from the Laboratoire de Physicochimie des Processus de Combustion et de l'Atmosphère (PC2A-CNRS) of Villeneuve d'Asq and C. Vovelle and J.L. Delfau from the Laboratoire de Combustion et Systèmes Réactifs (LCSR-CNRS) of Orléans from their useful advices in the experimental part.

**TABLE 1: REACTIONS OF ALLENE, PROPYNE, PROPARGYL AND DERIVED $C_4$ SPECIES**

The rate constants are given ($k=A\ T^n \exp(-E_a/RT)$) in cc, mol, s, kcal units. Reference numbers are given in brackets when they appear for the first time. The reactions in bold have been added or involve a modified rate constant compared to our last mechanism [8].

| Reactions | A | n | $E_a$ | References | No |
|---|---|---|---|---|---|
| ***Reactions of aC3H4 (CH2=C=CH2, allene)*** | | | | | |
| **aC3H4 = pC3H4** | **$2.5\times10^{12}$** | **0.0** | **59.0** | **Hidaka 89[27]** | **(1)** |
| aC3H4+M = C3H3+H+M | $2.0\times10^{18}$ | 0.0 | 80.0 | Hidaka 89 | (2) |
| **aC3H4+aC3H4 = C3H5+C3H3** | **$5.0\times10^{14}$** | **0.0** | **64.8** | **Dagaut92[28]** | **(3)** |
| C2H4+CH = aC3H4+H | $1.3\times10^{14}$ | 0.0 | -0.3 | Baulch 94[20] | (4) |
| C2H3+$^3$CH2 = aC3H4+H | $3.0\times10^{13}$ | 0.0 | 0.0 | Miller 92[3] | (5) |
| aC3H4+H = aC3H5 (high pressure limit) | $4.0\times10^{12}$ | 0.0 | 2.7 | Wagner 72[29] | **(6)** |
| **(low pressure limit)** | **$5.6\times10^{33}$** | **-5.0** | **4.44** | **Marinov96[4]** | |
| aC3H4+H = tC3H5 (high pressure limit) | $8.5\times10^{13}$ | 0.0 | 2.0 | Wagner 72 | **(7)** |
| **(low pressure limit)** | **$1.1\times10^{34}$** | **-5.0** | **4.44** | **Marinov96** | |
| iC4H3+$^3$CH2 = aC3H4+C2H | $2.0\times10^{13}$ | 0.0 | 0.0 | Miller 92 | (8) |
| **aC3H4+O => H+C2H3+CO** | **$6.6\times10^{12}$** | **0.0** | **3.0** | **Aleksandrov80[30]** | **(9)** |
| aC3H4+OH = CH2CO+CH3 | $2.0\times10^{12}$ | 0.0 | -0.2 | Liu 88 [31] | (10) |
| aC3H4+OH = HCHO+C2H3 | $2.0\times10^{12}$ | 0.0 | -0.2 | Liu 88 | (11) |
| **aC3H4+CH3 = C4H7-2** | **$8.0\times10^{12}$** | **0.0** | **7.4** | **Estimated[a]** | **(12)** |
| **aC3H4+ CH3 = iC4H7** | **$4.0\times10^{12}$** | **0.0** | **5.0** | **Estimated[b]** | **(13)** |
| **aC3H4+O2 = C3H3+HO2** | **$2.8\times10^{13}$** | **0.0** | **39.0** | **Estimated[c]** | **(14)** |
| **aC3H4+H = C3H3+H2** | **$1.3\times10^{6}$** | **2.53** | **9.2** | **Estimated[d]** | **(15)** |
| **aC3H4+O = C3H3+OH** | **$6.2\times10^{12}$** | **0.0** | **1.9** | **Estimated[e]** | **(16)** |
| **aC3H4+OH = C3H3+H2O** | **$6.2\times10^{6}$** | **2.0** | **0.4** | **Estimated[d]** | **(17)** |
| **aC3H4+CH3 = C3H3+CH4** | **$2.0\times10^{12}$** | **0.0** | **7.7** | **Wu87[32]** | **(18)** |
| **aC3H4+C2H = C3H3+C2H2** | **$1.0\times10^{13}$** | **0.0** | **0.0** | **Wu87** | **(19)** |
| **aC3H4+C2H5 = C3H3+C2H6** | **$5.0\times10^{14}$** | **0.0** | **19.8** | **Estimated[d]** | **(20)** |
| **aC3H4+C2H3 = C3H3+C2H4** | **$5.0\times10^{14}$** | **0.0** | **19.8** | **Estimated[d]** | **(21)** |
| **aC3H4+C3H5 = C3H3+C3H6** | **$2.0\times10^{12}$** | **0.0** | **7.7** | **Dagaut 92** | **(22)** |
| ***Reactions of pC3H4 (CH≡CCH3, propyne)*** | | | | | |
| pC3H4+M = C3H3+H+M | $4.7\times10^{18}$ | 0.0 | 80.0 | Hidaka 89 | (23) |
| C2H2+$^3$CH2 = pC3H4 | $3.5\times10^{12}$ | 0.0 | 0.0 | Tsang 86[20] | (24) |
| **pC3H4 = C2H+CH3** | **$4.2\times10^{15}$** | **0.0** | **125.0** | **Estimated[f]** | **(25)** |
| **pC3H4+H = C2H2+CH3** | **$1.3\times10^{5}$** | **2.5** | **1.0** | **Hidaka 89** | **(26)** |
| pC3H4+H = tC3H5 (high pressure limit) | 8.5E12 | 0.0 | 1.7 | Wagner 72 | **(27)** |
| **(low pressure limit)** | **$5.6\times10^{25}$** | **-7.27** | **6.58** | **Marinov96** | |
| pC3H4+H = sC3H5 (high pressure limit) | 5.8E12 | 0.0 | 3.1 | Wagner 72 | **(28)** |
| **(low pressure limit)** | **$3.8\times10^{25}$** | **-7.27** | **7.98** | **Estimated[g]** | |
| pC3H4+O = CHCO+CH3 | $1.5\times10^{13}$ | 0.0 | 2.1 | Warnatz 84 [33] | (29) |
| pC3H4+OH = CH2CO+CH3 | $4.3\times10^{11}$ | 0.0 | -0.8 | Boodaghians 87 [34] | (30) |
| **pC3H4+OH =H+C2H3CHO** | **$4.3\times10^{11}$** | **0.0** | **-0.8** | **Estimated[h]** | **(31)** |
| pC3H4+HO2 = C2H4+CO+OH | $6.0\times10^{9}$ | 0.0 | 8.0 | Estimated[i] | (32) |
| **pC3H4+O2 = C3H3+HO2** | **$2.1\times10^{12}$** | **0.0** | **40.8** | **Estimated[c]** | **(33)** |
| pC3H4+H = C3H3+H2 | $1.7\times10^{5}$ | 2.5 | 2.5 | Estimated[j] | (34) |
| pC3H4+CH3 = C3H3+CH4 | 2.2 | 3.5 | 5.7 | Estimated[j] | (35) |
| pC3H4+C2H = C3H3+C2H2 | $3.6\times10^{12}$ | 0.0 | 0.0 | Estimated[j] | (36) |
| pC3H4+C2H3 = C3H3+C2H4 | 2.2 | 3.5 | 4.7 | Estimated[j] | (37) |
| pC3H4+C2H5 = C3H3+C2H6 | 2.2 | 3.5 | 6.6 | Estimated[j] | (38) |
| **pC3H4+O = OH+C3H3** | **$3.4\times10^{4}$** | **2.16** | **4.8** | **Adusei96[35]** | **(39)** |
| pC3H4+OH = C3H3+H2O | $3.1\times10^{6}$ | 2.0 | -0.3 | Estimated[j] | (40) |
| pC3H4+HO2 = C3H3+H2O2 | $9.6\times10^{3}$ | 2.6 | 13.9 | Estimated[j] | (41) |

***Reactions of C3H3 (CH≡CCH2•, resonance stabilized propargyl radicals)***

| | | | | | |
|---|---|---|---|---|---|
| **C3H2+H = C3H3** | **1.0E14** | **0.0** | **0.0** | **Estimated[k]** | **(42)** |
| [1]CH2+C2H2 = C3H3+H | 1.8E14 | 0.0 | 0.0 | Fournet99[8] | (43) |
| C3H3+H = C3H2+H2 | 2.0E13 | 0.0 | 0.0 | Fournet99 | (44) |
| C3H3+O = C2H+HCHO | 1.4E14 | 0.0 | 0.0 | Fournet99 | (45) |
| C3H3+OH = C3H2+H2O | 2.0E13 | 0.0 | 0.0 | Miller 92 | (46) |
| C3H3+OH = C2H3+CHO | 4.0E13 | 0.0 | 0.0 | Wang 97 [5] | (47) |
| C2H2+CHCO = C3H3+CO | 1.0E11 | 0.0 | 3.0 | Miller 92 | (48) |
| C3H3+O2 = CH2CO+CHO | 3.0E10 | 0.0 | 2.9 | Milller 92 | (49) |
| **C3H3+OOH => OH+C2H+HCHO** | **$1.0x10^{15}$** | **-0.8** | **0.0** | **Estimatedl[l]** | **(50)** |
| C3H3+C3H3 = C6H5+H | 1.0E12 | 0.0 | 0.0 | Stein 90[36] | (51) |

[a] : Rate constant of this addition of methyl radicals considered as twice the value proposed by Tsang [37] for propene assuming a fall-of effect of a factor 4 to take into account the experimental pressure of 6.7 kPa.

[b] : Rate constant of this addition of methyl radicals taken equal to the value measured by Tsang [38] at pressures between 1.5 and 5 bar assuming the same fall-of effect as for reaction (12).

[c] : Rate constant of the bimolecular initiation with oxygen molecule calculated as proposed by Ingham et al [39] (see text).

[d] : Rate constant estimated as that of the similar reaction in the case of 1,3-butadiene [8].

[e] : Rate constant for the H-abstraction calculated at 1000 K from the values proposed by Aleksandrov *et al.* [30] for rate constants of the global reaction and of the addition.

[f] : Rate constant of this unimolecular initiation calculated by the modified collision theory at 1500 K using software KINGAS [40].

[g] : Rate constant at low pressure limit deduced from the rate constant at low pressure limit proposed by Marinov *et al.* [4] for reaction (27) and from the ratio between the rate constants at high pressure limit of reactions (27) and (28).

[h] : Rate constant of this addition taken equal to that of reaction (30)

[i] : Rate constant estimated as that of the similar reaction in the case of acetylene [20].

[j]: Rate constant estimated as that of the similar reaction in the case of propene as proposed by Tsang [37].

[k]: Rate constant taken equal to that of the recombination of •H atoms with allyl radicals as proposed by Allara *et al.* [41].

[l]: Rate constant estimated according to the correlation proposed for allyl radicals by Heyberger *et al.* [26].

## FIGURE CAPTIONS

Figure 1: Typical chromatogram of $C_3$-$C_6$ compounds obtained for the flame doped with allene at a distance of 0.48 cm from the burner (oven temperature program: 313 K during 22 min, then a rise of 1 K/min until 523 K).

Figure 2: Temperature profiles in the three flames: (a) experimental measurements performed without the sampling probe, (b) profiles used for simulation compared in the case of allene to the experimental profiles obtained with and without the probe.

Figure 3: Profiles of the mole fractions of oxygen and $C_1$ species in the three flames. Points are experiments and lines simulations. Stars and thin lines correspond to the flame of pure methane, white circles and dotted lines to the flame doped with allene and black circles and full lines to the flame seeded with propyne. The broken line is related to a simulated flame of pure methane at $\Phi$=1.55 (see text), but cannot be seen in fig. 3a.

Figure 4: Profiles of the mole fractions of hydrogen and $C_2$ species in the three flames. Points are experiments and lines simulations. Stars and thin lines correspond to the flame of pure methane, white circles and dotted lines to the flame doped with allene and black circles and full lines to the flame seeded with propyne. The broken line is related to a simulated flame of pure methane at $\Phi$=1.55 (see text).

Figure 5: Profiles of the mole fractions $C_3$ species and benzene in the three flames. Points are experiments and lines simulations. Stars and thin lines correspond to the flame of pure methane, white circles and dotted lines to the flame doped with allene and black circles and full lines to the flame seeded with propyne.

Figure 6: Profiles of the mole fractions $C_4$ species in the two doped flames. Points are experiments and lines simulations. White circles and dotted lines correspond to the flame doped with allene and black circles and full lines to the flame seeded with propyne.

Figure 7: Flow rate analysis for the consumption of the $C_3$ reactant in (a) the flame seeded with allene and (b) the flame seeded with propyne for a distance of 0.12 cm from the burner corresponding to a temperature of 720 K and a conversion of 35 % of the $C_3$ reactant.

Figure 8: Simulated profiles of (a) propargyl radicals and (b) hydrogen atoms in the doped flames.

Figure 9: Flow rate analysis for the formation and consumption of phenyl radicals in the flame seeded with allene for a distance of 0.25 cm from the burner corresponding to a temperature of 1290 K, a conversion of 75% of the $C_3$ reactant and to the peak of benzene profile.

Figure 1

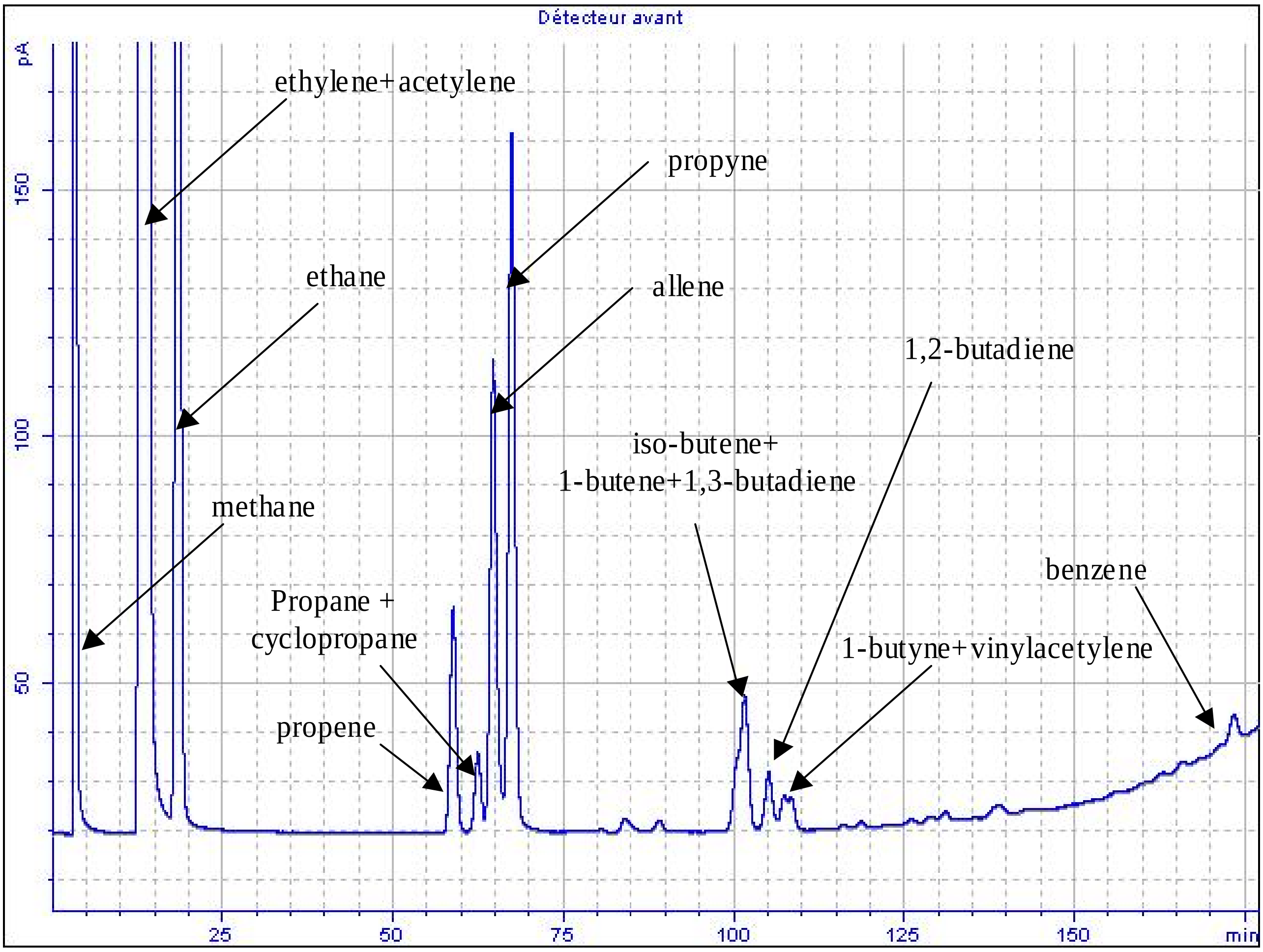

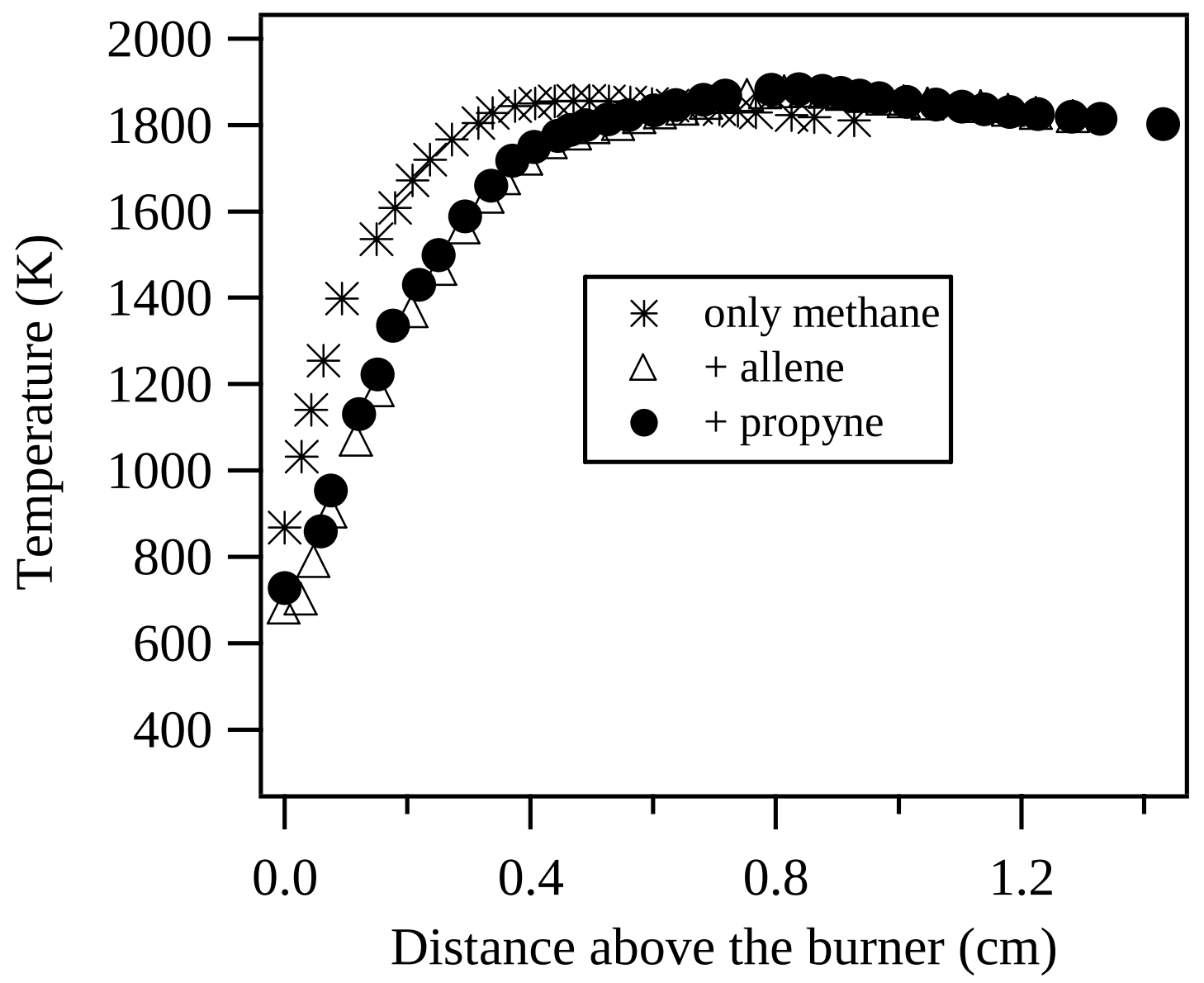

(a)

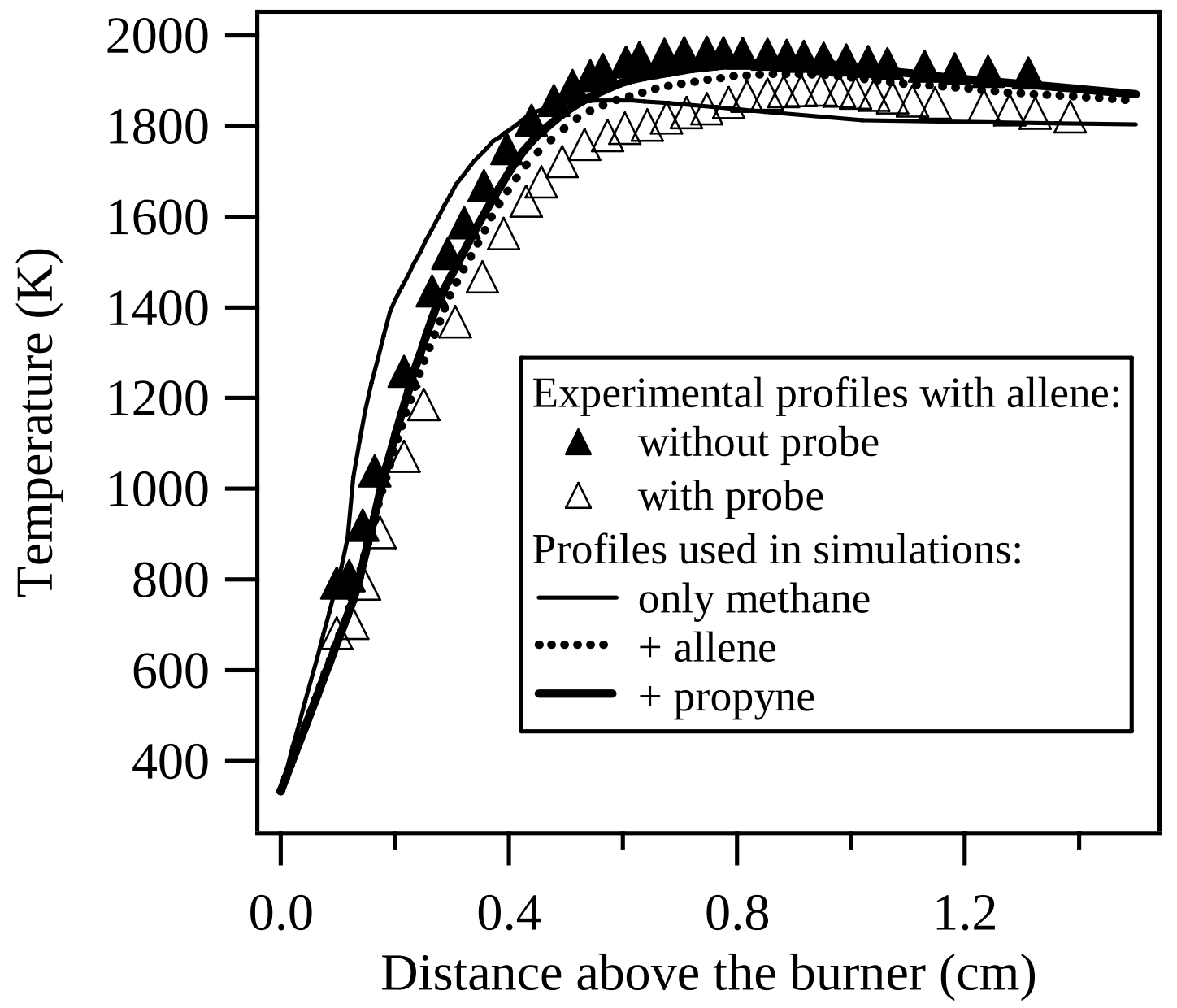

(b)

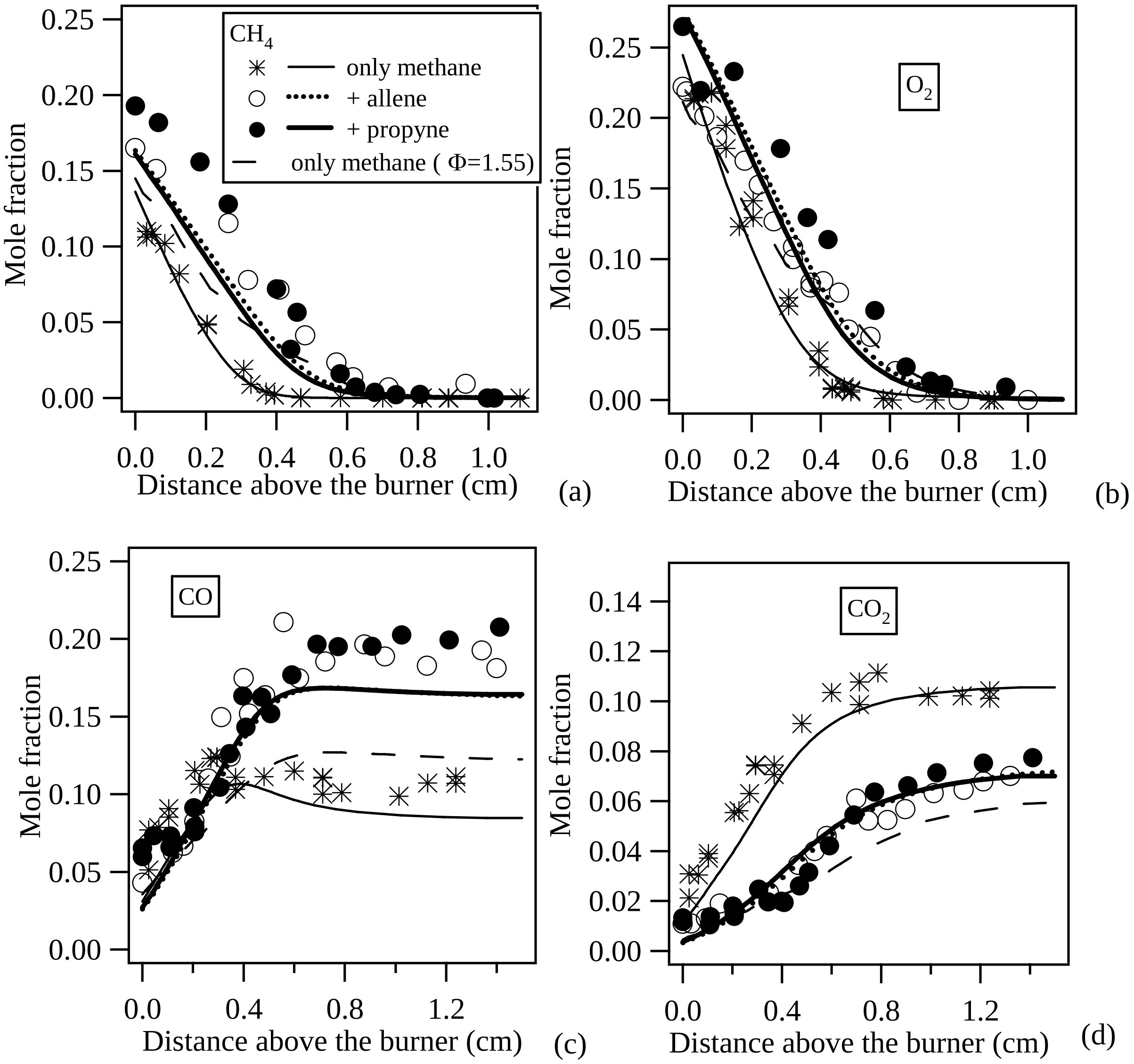

Figure 4

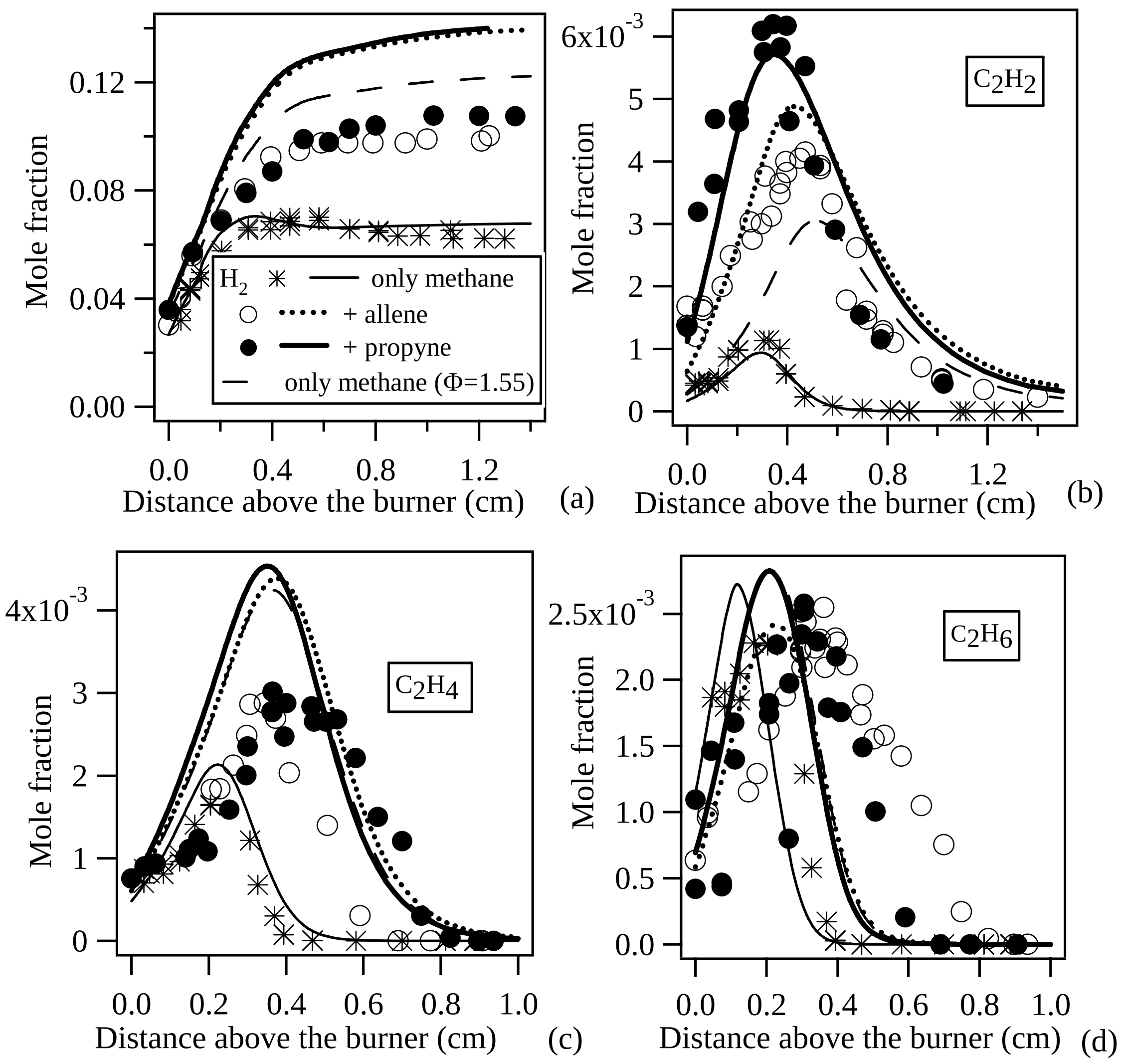

Figure 5

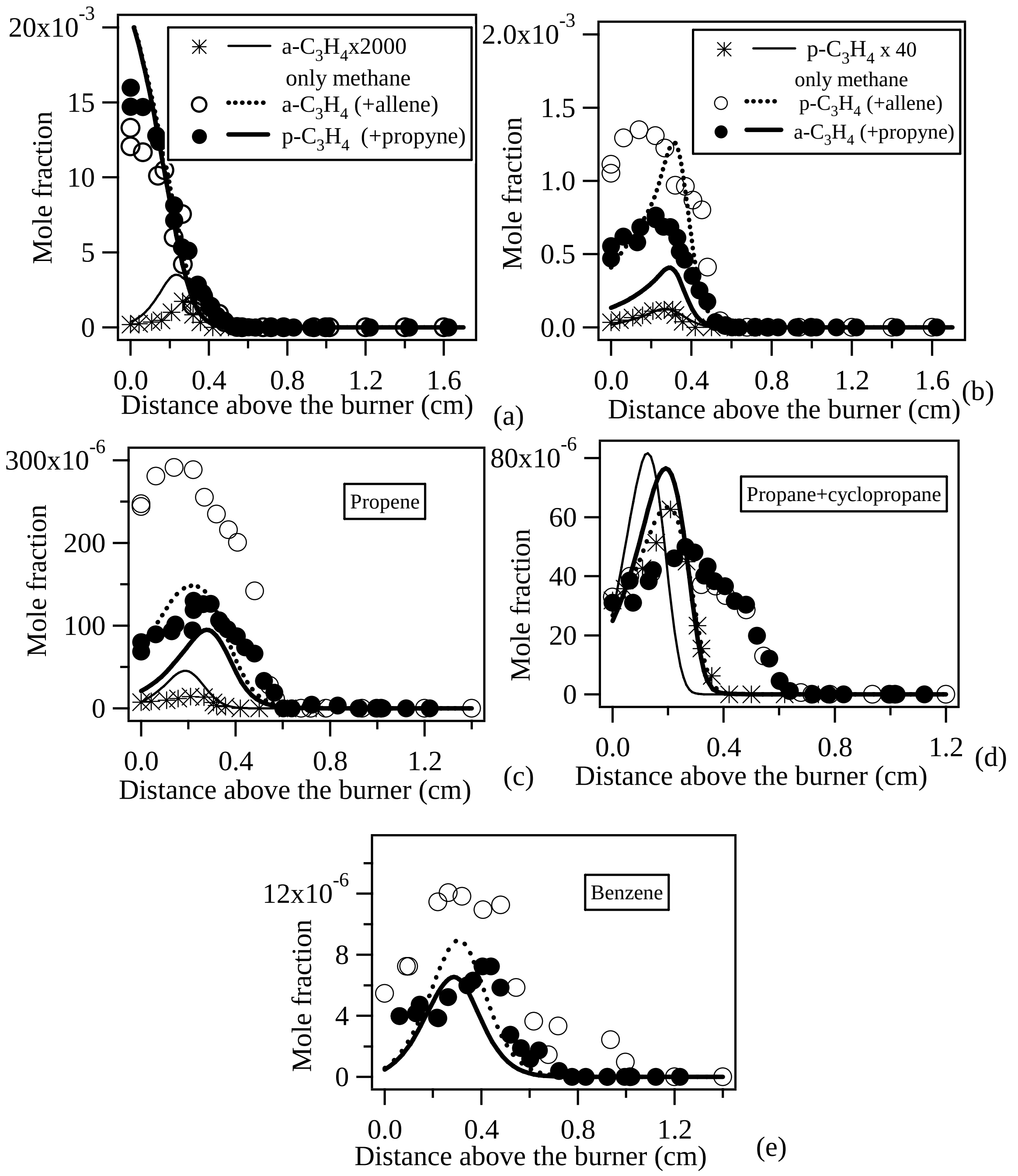

Figure 6

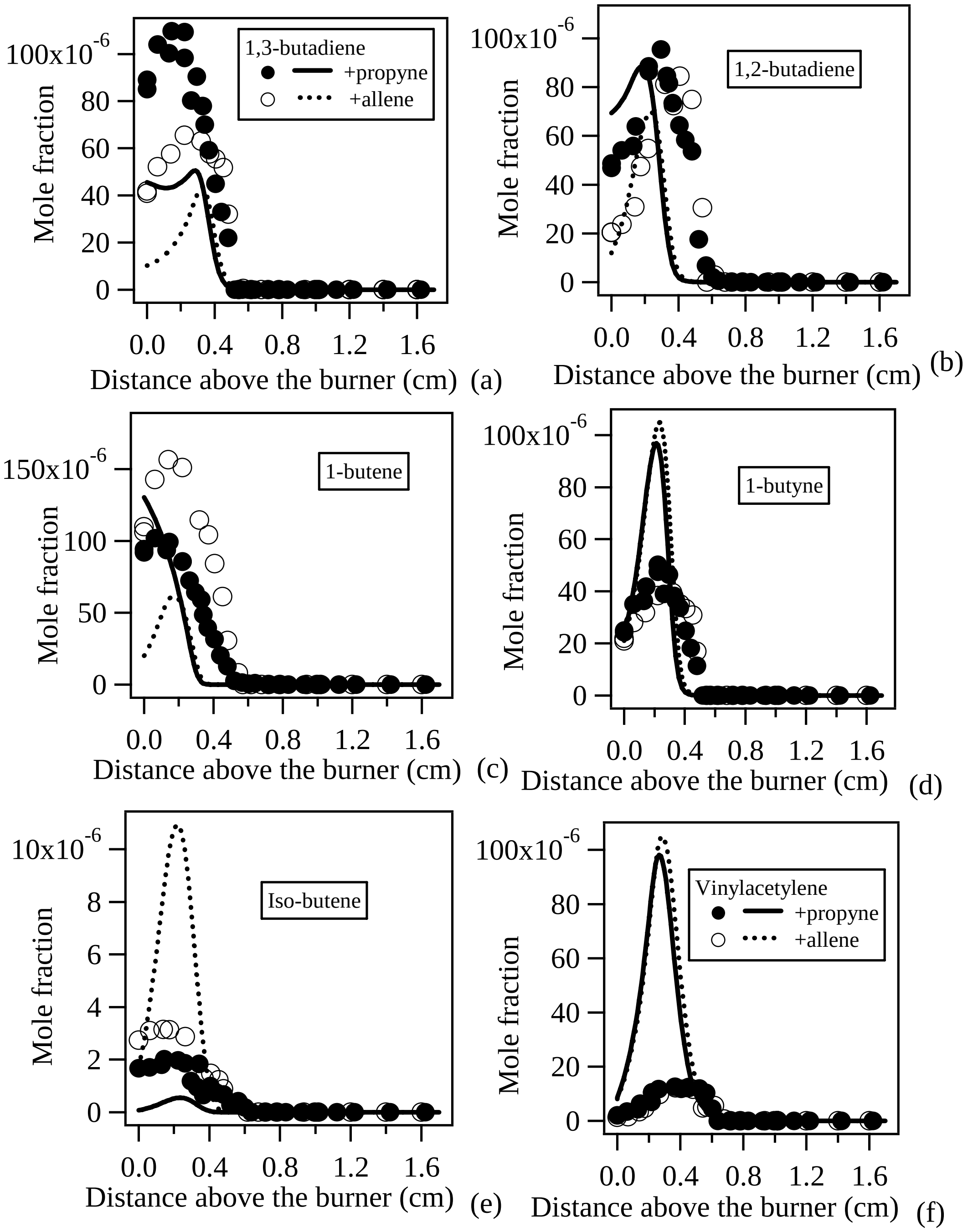

Figure 7

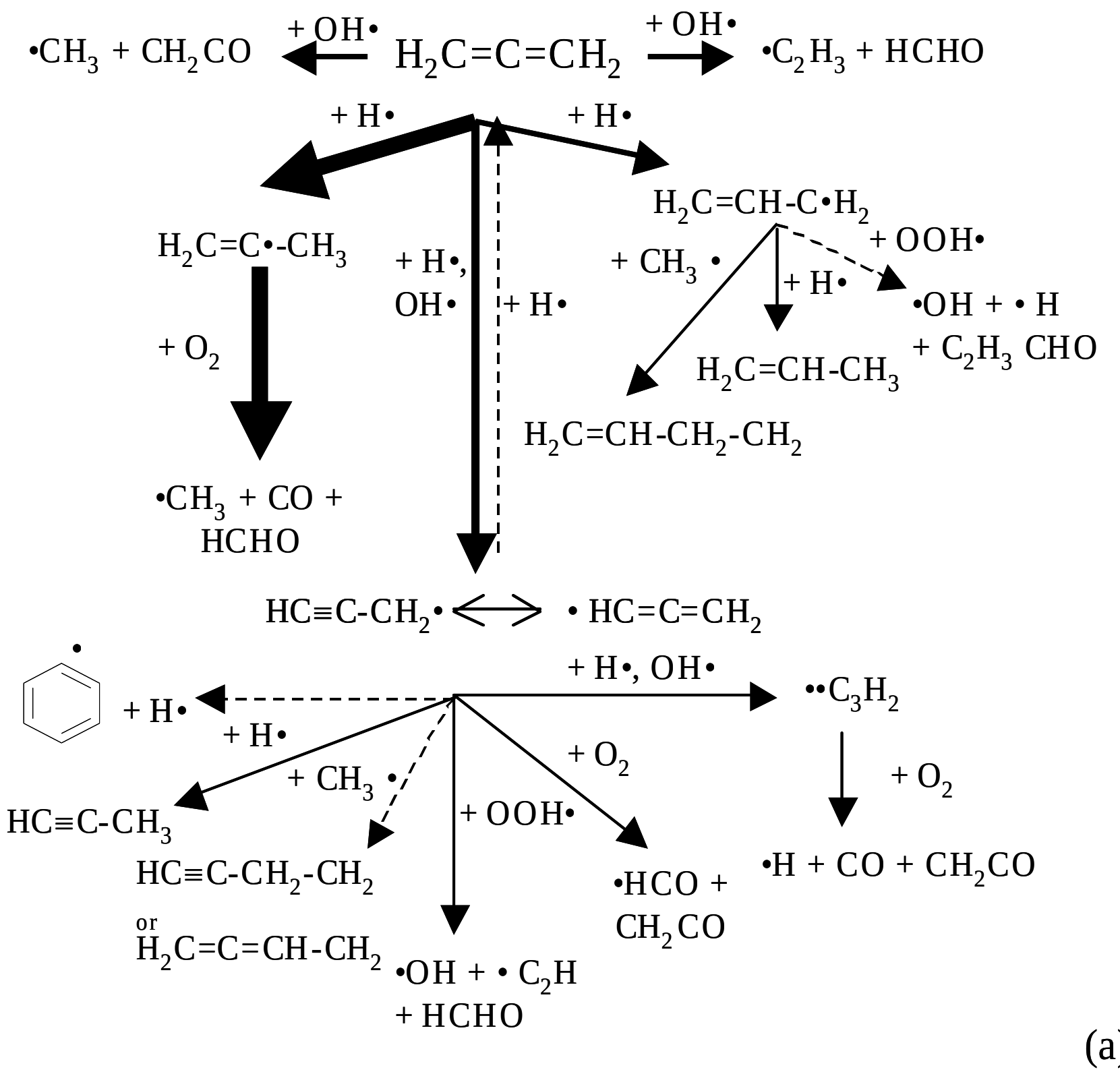

(a)

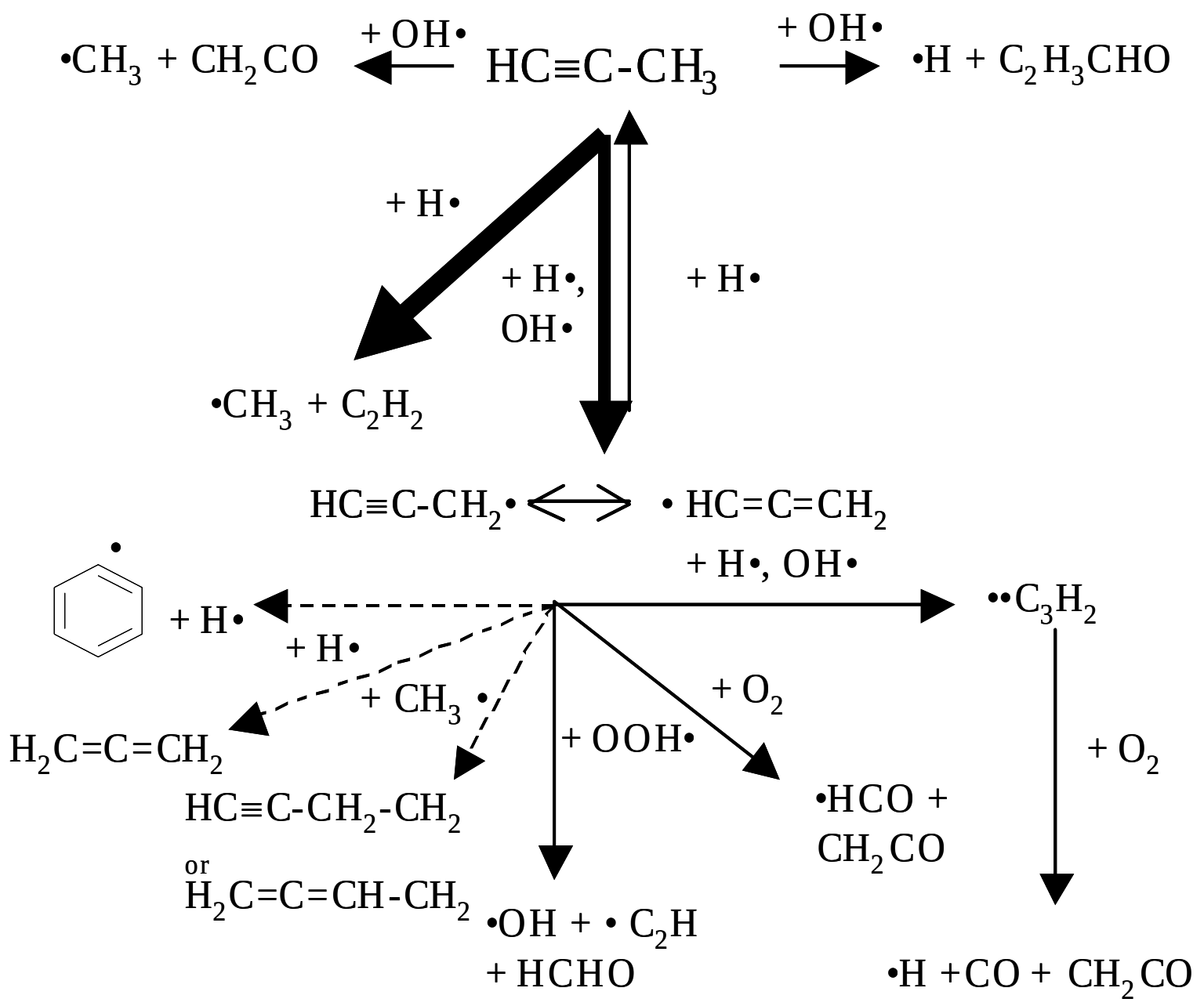

(b)

Figure 8

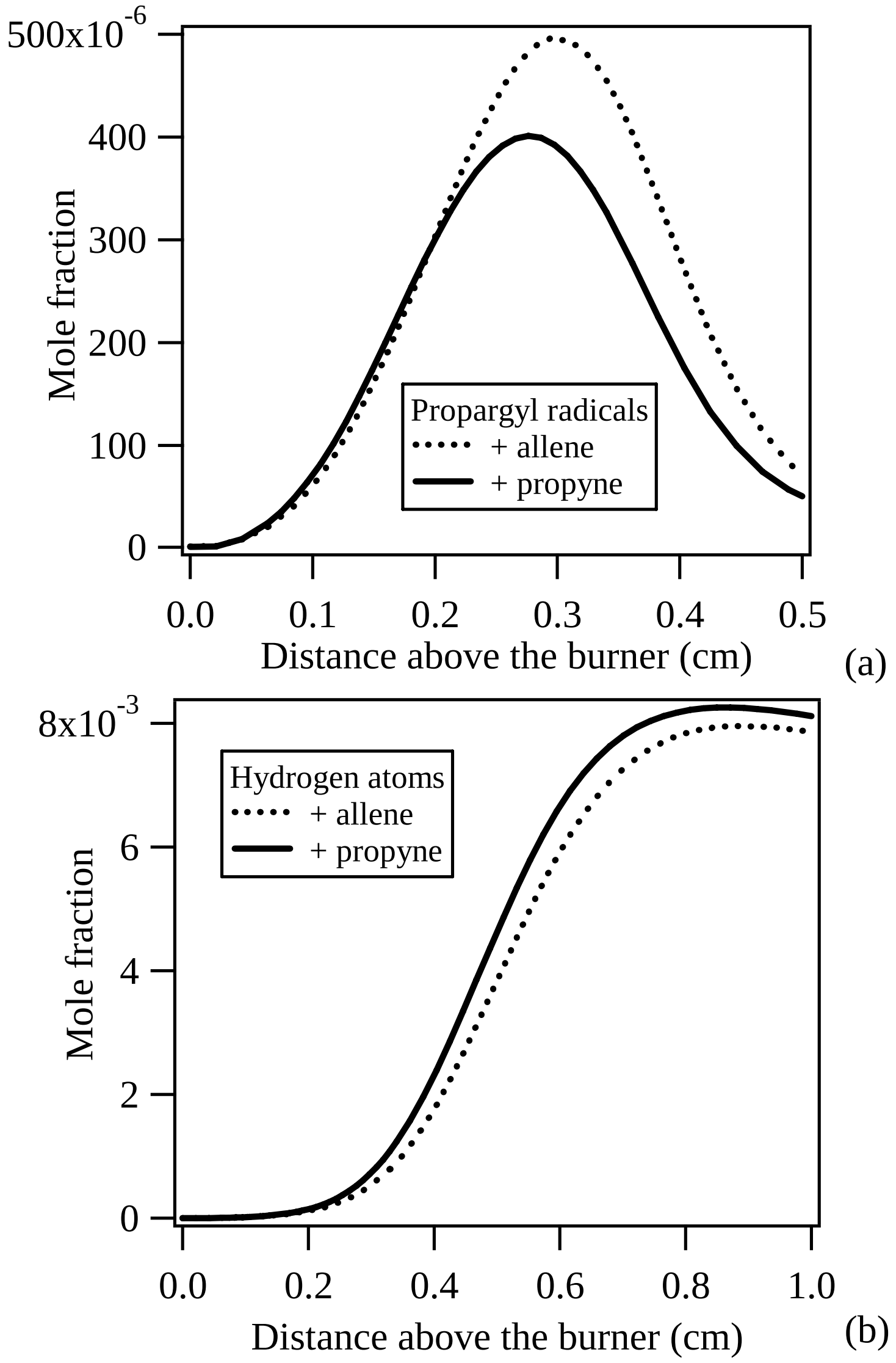

Figure 9

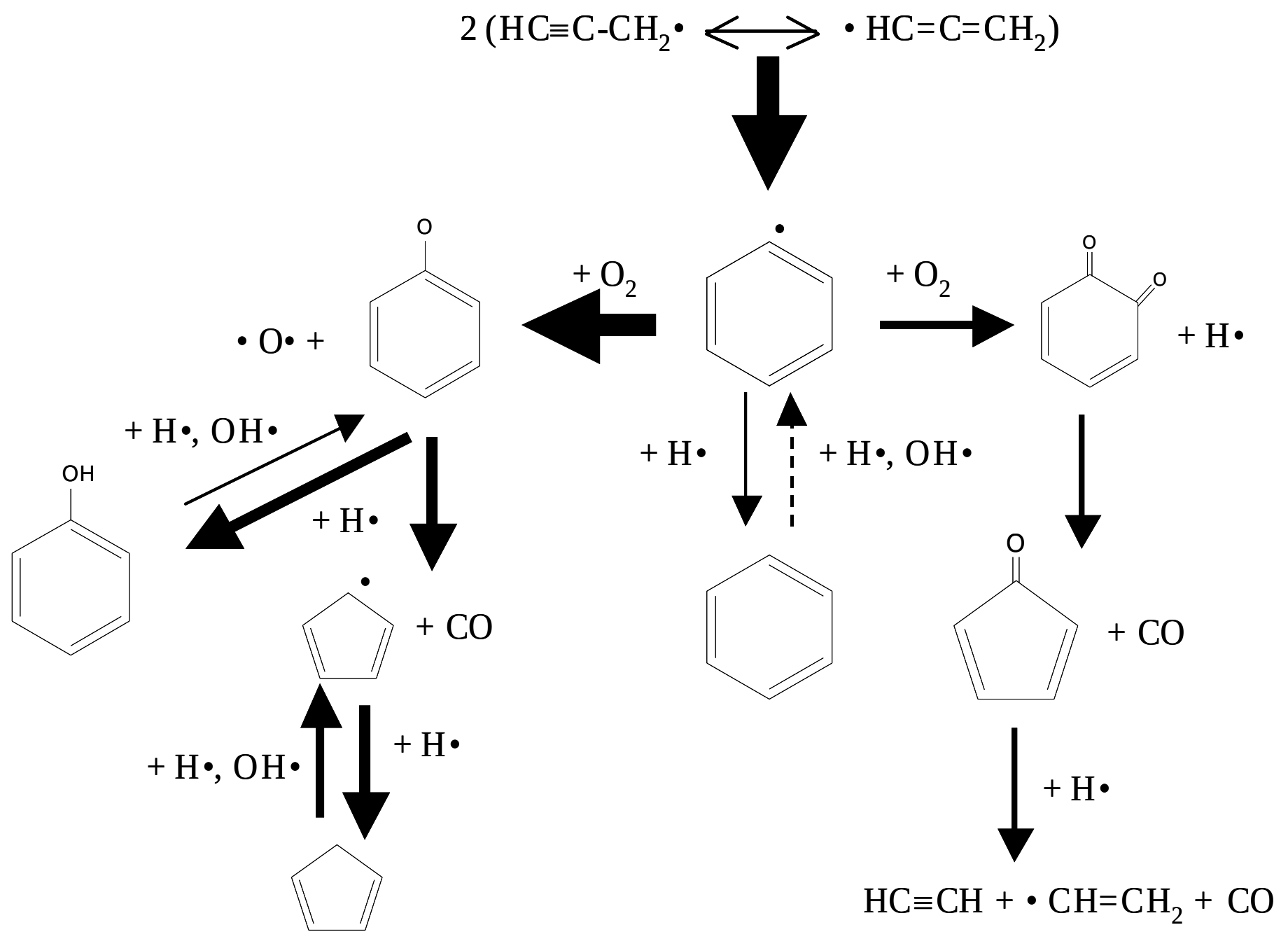